\documentstyle[twocolumn,aps,psfig]{revtex}
\begin{document}
\newcommand{\beq}{\begin{equation}}
\newcommand{\eeq}{\end{equation}}

\title{Hydrophobicity and Unique Folding of Selected Polymers}
\author{Michele Vendruscolo}
\address{Department of Physics of Complex Systems, Weizmann Institute of Science
, Rehovot 76100, Israel}

\address{
\centering{
\medskip\em
{}~\\
\begin{minipage}{14cm}
{}~~~
In suitable environments, proteins, nucleic acids and certain
synthetic polymers fold into unique conformations.
This work shows that it is possible to construct lattice
models of foldable heteropolymers by expressing the energy only in terms
of individual properties of monomers, such as
the exposure to the solvent and the steric factor.
{}~\\
{}~\\
{\noindent PACS numbers: 87.15.By, 87.10.+e }
\end{minipage}
}}
\maketitle

It is generally believed that the hydrophobic interaction plays a major role 
in protein folding 
\cite{creighton,dill,bw87,jones96,got,pgt97,s97,ct93,tang,irb96}.
Under physiological conditions,
non polar amino acids are buried inside the core of the native state of a
protein to avoid contact with water molecules.
A long standing question is to what extent other non-covalent forces,
such as hydrogen bonding, electrostatic and van der Waals interactions 
contribute to stabilize the folded 
state \cite{creighton,dill,bw87,jones96}. 

Unraveling the different roles played by these interactions
will have a considerable impact in different areas of research in biophysics,
such as the prediction 
of protein structures \cite{creighton,dill,got,pgt97,s97}
the design of synthetic drugs \cite{pgt97,sg93,deutsch,seno},
and the production of self-assembling non-biological polymers \cite{oligomers} 
and other polymeric materials \cite{muth}.

With the advent of genome projects \cite{duboule}
a wide gap is opening between the number of known protein sequences 
and their correspondent structures \cite{sussman}.
The bottleneck in protein structure prediction is at present largely due
to the incorrect treatment
of the interactions \cite{pgt95,f97,stab}.
The current state of the art 
is highlighted by a recent study by Fisher and Eisenberg \cite{fisher97}. 
They carried out the assignment of structures to the sequences
encoded in the complete genome of {\em Mycoplasma genitalium}.
Among the complete set of 468
sequences, they were able to assign 103 (22\%) of them to
a structure with high confidence.
They used homology modelling and threading techniques that are at present
the most successful prediction tools available. 
Homology modelling is based on the observation
that the exact identity of amino acids is not crucial
for maintaining the overall fold of a protein \cite{fisher96}.
Proteins that differ as much as 70\% in their sequences usually
share the same fold.
Thus, it is possible to predict the conformation of a sequence by using
a set of experimentally determined protein structures 
with similar amino acid composition.
Threading relies on the surprising fact that more distantly related proteins,
whose sequence similarity is close to the threshold 
of pure randomness, do sometimes share the same fold \cite{jones96}.
The search for {\em compatibility} between sequence and structure
has inspired various techniques 
to single out the native state of a protein from
a library of alternative structures
\cite{jones96,fisher96,bowie91}.
The screening is typically carried out by assigning an energy-like function
that incorporates the compatibility of each amino acid 
to its local environment. 
Compatibility is described in terms of charge, polarity and 
secondary structure content, within a given conformation.
Details in the local environment play a crucial role also in RNA folding.
A key ingredient in this case is given by metal ion
coordination numbers \cite{rna}.
Likewise and rather surprisingly, a non-biological polymer 
(an aromatic hydrocarbon) has been recently designed which is able
to fold into a unique helical structure having a large cavity, supposedly
under the effect of the hydrophobic interaction \cite{oligomers}.

This letter contributes to the development of a rational treatment 
of the hydrophobic interaction at the single monomer level. 
We show that it is possible to construct minimalistic lattice models 
of foldable heteropolymers, by introducing an energy function that depends 
only on individual residues' environments.
A model will be called ``foldable'' if there are sequences,
either randomly chosen or selected,
with a unique, thermodynamically stable and kinetically reachable ground state
\cite{dill,bw87,pgt97,s97,ct93,tang,stab,ssk94,bosw95,thir97}.
We adopt a simple approximation for the energy
which accounts both for the propensity to be exposed to the solvent 
and for the excluded volume effects due to the
different sizes of the monomers.
Although naturally existing or synthesized polymers,
such as proteins, nucleic acids and tailored hydrocarbons,
are characterized by much more complex interactions, 
the main focus here is on the fact that the unifying feature
is the tendency to avoid contact with the solvent 
by some species of monomers.
Previous theoretical studies concentrated mainly on the treatment
of pairwise contact energies \cite{dill,pgt95,stab,mj96,glw92,hl94,ms96,md96}.
This is in contrast with the present study, in which the hydrophobic effect 
is investigated at the individual particle level.

Lattice models, although often criticized \cite{honig}, have been
recognized to capture some of the most relevant thermodynamic features of the
folding process \cite{s96}, such as the existence of a unique
ground state, amenable to exact computations, and the cooperativity of the
transition. Even key dynamical processes, such as the nucleation-condensation
mechanism \cite{fersht}, have been validated with the help of lattice models
\cite{sap96}.

Results obtained on a 2D square lattice are presented first.
On a lattice, a polymer is represented as a connected chain of $N$ monomers.
Hydrophobicity and steric factors
can be modeled as the tendency of a monomer to have a specific number 
of non-bonded nearest-neighbors.
We define the hydrophobic model HM$_1$ by expressing
the energy $E_1$ as,
\begin{equation}
E_1 = \sum_{i=1}^N | n({a_i}) - {\overline n}(a_i) | \; ,
\label{eq:en1}
\end{equation}
where $n(a_i)$ is the number of non-bonded nearest-neighbors
of monomer of species $a_i$ in position $i$ along the chain, 
and ${\overline n}(a_i)$ is the ideal value of $n(a_i)$.
This expression was first proposed by Hao and Scheraga, who
supplemented it to a pairwise energy term \cite{hs97}.
They presented a method to optimize energy parameters to obtain
lattice models of foldable polymers.
Other previous work has been devoted to 
the hydrophobic interaction, although
without specifically disentangling it from other interactions.
Mirny and Domany \cite{md96} introduced explicitly an hydrophobic term
in the energy function and they performed various tests of fold recognition
and dynamics.
In a recent work Li, Tang and Wingreen \cite{hpli}
discussed the ``designability principle'' \cite{tang}
in terms of a ``binary'' model with two species of amino acids, where the
energy is expressed in terms of the exposure to the solvent only.
The model proposed here is much more general and no major modification
is required to extend it to the treatment 
of realistic models of foldable heteropolymers,
as for example, in a ``contact map'' representation of protein structure
\cite{md96,vkd97}.

We will first investigate if in 2D the HM$_1$ model gives rise to foldable
sequences and we will compare our results with those obtained using
the standard HP model \cite{dill}.
In 2D, with the aid of the HP model \cite{dill,ct93,tang,seno},
it has been demonstrated that 
it suffices to assume only 2 species of monomers
to guarantee uniqueness of the ground state, although probably
not the right order of the folding transition \cite{s97}.
In the HP model the energy is written in the pairwise contact approximation
\begin{equation}
E_{pair}= \sum_{j>i} U(a_i,a_j) \Delta_{ij} \;,
\label{eq:pair}
\end{equation}
where $a_i$ can be either $H$ (hydrophobic) or $P$ (polar) and
$\Delta_{ij}$ is a contact matrix, which is defined to be 
1 if two monomers are non-bonded nearest-neighbor and 0 otherwise.
The typical values for the interaction parameters 
are $U(H,H)=-1$ and $U(H,P)=U(P,P)=0$ \cite{dill}.
A chain of $N=16$ monomers is amenable to complete enumeration
of all 802075 possible symmetry-unrelated
conformations, either compact or not \cite{dill,deutsch,seno}.
For the above mentioned choice of contact energy parameters, 
there are 1539 (2\%) sequences among 
the $2^{16}=65536$ possible ones which have a unique ground 
state \cite{deutsch,seno}.
We compare this result with those obtained by using Eq.(\ref{eq:en1}),
setting ${\overline n}(1)=1.5$ (hydrophobic-like) 
and ${\overline n}(2)=0.4$ (polar-like). 
A larger number, $10178$ (16\%), of sequences was found to
have a unique ground state.

We have also explored the case of 
three species of monomers, a number which, within a contact
approximation of the interactions (as in Eq.(\ref{eq:pair})), is believed
to epitomize the essential features of the interplay between folding
and glass transitions in random heteropolymers \cite{owls95}.
We chose at random 20177 sequences
among the 2018016 possible ones with fixed composition 
$N(1)=6$, $N(2)=N(3)=5$, where $N({a})$
is the number of monomers of species $a$.
Choosing energy parameters ${\overline n}(1)=0.4$,  
${\overline n}(2)=1.1$ and ${\overline n}(3)=1.8$, 
9439 (47\%) sequences were found with a unique ground state.

In the spirit of Mirny and Domany \cite{md96},
a more realistic form for the hydrophobic energy is given by 
\begin{equation}
E_2 = \sum_{i=1}^N 
\beta(a_i)\left[ n(a_i) - {\overline n}(a_i) \right]^2 \; .
\label{eq:en2}
\end{equation}
This expression will be referred to as hydrophobic model HM$_2$.
The parameters $\beta(a_i)$ capture the various degrees
with which the different species of monomers tend to attain
the preferred number ${\overline n}(a_i)$ of contacts.
In the case of 2 species of monomers, we repeated the same calculation 
as for the HM$_1$ model.
Letting here $\beta(1)=\beta(2)=1$, we found $9821$ (14\%) sequences 
with a unique ground state.
We observe that, at least from the above calculations in 2D, the
approximation of the hydrophobic interaction
proposed in this work is capable of yielding foldable sequences.

We now turn to the calculations in 3D which represent the essential part
of this work, and further illustrate the extent to which the present model 
embodies foldability.
It is known that the HP model is pathological in 3D, since it is
rather uncommon to have a sequence with a unique ground state 
with a large gap above it \cite{s97,6packs},
although the situation can be different 
with a choice of parameters favoring 
more collapsed structures \cite{tang,stab}.
We discuss here the general case of 20 species of monomers in the HM$_2$ model.
We compare these results with 
those obtained by using a common parametrization
of the pairwise contact interaction matrix $U(a_i,a_j)$ in Eq.(\ref{eq:pair}),
due to Miyazawa and Jernigan (MJ) \cite{mj96}, 
although other choices would be possible \cite{mj96,glw92,hl94,ms96,md96}.
For the HM$_2$ model,
we derived the 40 parameters ${\overline n}(a)$ and $\beta(a)$ 
from a statistical analysis of the non redundant
set of 246 protein structures reported by Hinds and Levitt \cite{hl94}.
The procedure, similar to that of Mirny and Domany \cite{md96},
is straightforward. For each amino acid species $a$, we computed 
the average, ${\overline n}(a)$,
and the standard deviation, $\beta(a)$, of the number of contacts it
forms in the set of experimentally known crystal structures
(see Table \ref{tab:anu}). Two amino acids are said to be in contact 
if their $C_{\alpha}$ atoms are closer 
than 8.5 \AA \hspace{2pt} in the native structure \cite{vkd97}.

On the cubic lattice, the 103346 symmetry-unrelated maximally
compact conformations of a polymer
of length $N=27$ can be enumerated in a manageable computer time 
\cite{tang,sg90}.
If it is guaranteed that the ground state is maximally compact,
exact enumeration can be used to demonstrate its uniqueness.
We adapted the energy parameters to the cubic lattice by 
matching the average number of contacts
that a monomer forms on the $3\times 3\times 3$ cube with
the average of the ideal number of contacts, 
$(1/20)\sum_{a=1,20} {\overline n}(a)$.
This result is obtained by rescaling the energy parameters
in Table \ref{tab:anu} by a factor 3.315. 

\begin{figure}
\centerline{\psfig{figure=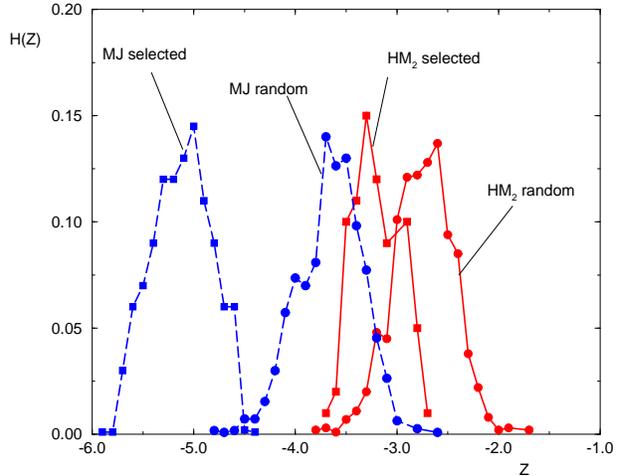,height=7.5cm,angle=270}}
\caption{Normalized histograms of the $Z$ scores for 
the HM$_2$ (full lines) and MJ (dotted lines) models.
Circles refer to random sequences and squares to designed ones.}
\label{fig:z}
\end{figure}

To characterize foldability, we first investigate the thermodynamic stability
of the ground states of random sequences. 
A typical measure of thermodynamic stability 
is given by the $Z$ score \cite{stab,ms96}, which is defined by
$Z= ({E_n - \langle E \rangle})/{\sigma} $,
where $E_n$ is the energy of the ground state, $\langle E \rangle$
is the average energy, and $\sigma$ the standard deviation in the
distribution of the energy around the average.
We measure the distribution of the $Z$ scores for a set of 1000
random HM$_2$ sequences. 
We found that only 2\% of them had a unique
lowest energy state (the ``ground state''
among maximally compact conformations). Moreover,
on average the degeneracy was 22.
For comparison, 99\% of the 1000 random MJ sequences that we considered
had a non degenerate lowest energy state, 
and the remaining ones had a very small degeneracy.
The result of the comparison of the $Z$ scores is 
shown in Fig. \ref{fig:z}.

The low values of the $Z$ score and the large degeneracy
found for the HM$_2$ case mark
a shortcoming of enumerating only maximally compact conformations.
By using other simulation techniques, 
such as the standard lattice Monte Carlo (SMC) \cite{ssk94,so}, 
and the prune-enriched Rosenbluth method (PERM) \cite{grass}, 
we easily found non-compact lower energy states for most
of the considered HM$_2$ sequences.
It is known, however, that foldability is a property of {\em selected}
sequences \cite{got,pgt97,s97,irb96,sg93,deutsch,seno,bosw95,thir97}.
A way to demonstrate that the HM$_2$ model is foldable is to show that 
it possible to select sequences whose ground states are both
unique {\em and} maximally compact.
The usual design procedure \cite{sg93}, 
introduced to study pairwise interactions, prescribes to 
choose a target conformation and then to search in sequence space
for the sequence with minimal energy onto such conformations.
This procedure delivers a better $Z$ score for the 1000 designed MJ sequences
that we considered, as can be seen from Fig. \ref{fig:z}.
However, in the case of the HM$_2$ model, 
we found that such technique is not sufficiently
effective in designing out alternative conformations.
A sequence design procedure similar to those
proposed in Refs. \cite{deutsch,seno} proved to be more effective.
Sequences selected in this way were found to have a unique ground state
by exact enumeration among maximally compact conformations.
More crucially, in no cases we have been able to reach lower energy states
using dynamical simulation techniques such as the SMC and the PERM algorithms.
The histogram of the $Z$ score of the
100 HM$_2$ sequences selected in this way is shown in Fig. \ref{fig:z}.

In summary, we have shown both in 2D and in 3D that
it is possible to construct simple models of foldable heteropolymers 
by expressing the hydrophobic and the steric interactions 
at the level of individual monomers.

It is a pleasure to thank E. Domany and P. Grassberger for discussions.



\newpage
\begin{minipage}{17cm}
\begin{table}
\begin{center}
\begin{tabular}{|c|c|c|c|c|c|c|c|c|c|c|c|c|c|c|c|c|c|c|c|c|}
ALA & GLU & GLN & ASP & ASN & LEU & GLY & LYS & SER & VAL & ARG & THR & PRO & ILE & MET & PHE & TYR & CYS & TRP & HIS \\
\hline
 7.56 & 5.62 & 6.23 & 5.51 & 6.02 & 7.63 & 5.55 & 5.86 & 6.31 & 8.29 & 6.58 & 6.73 & 5.73 & 8.07 & 7.72 & 7.58 & 7.45 & 8.81 & 7.67 & 6.59 \\
\hline
  2.98 & 2.17 & 2.36 & 2.41 & 2.58 & 2.25 & 3.51 & 2.08 & 3.07 & 2.53 & 2.33 & 2.77 & 2.85 & 2.35 & 2.37 & 2.31 & 2.51 & 2.42 & 1.36 & 2.43 \\
\end{tabular}
\vspace{0.5cm}
\caption{Mean ${\overline n}(a)$ 
and standard deviation $\beta(a)$ of the number of contacts
of amino acids, obtained from a statistical analysis of a non redundant set
of 246 protein structures.}
\label{tab:anu}
\end{center}
\end{table}
\end{minipage}

\end{document}